\documentclass{article}
\def\today{16.6.14} 
\usepackage{amsmath,amsfonts,amsthm,amssymb,amscd}
\usepackage{graphicx}
\theoremstyle{plain} 
\newtheorem{theorem}{Theorem}[section]

\newtheorem{corollary}[theorem]{Corollary} 
\newtheorem{remark}[theorem]{Remark}

\newcommand{\R}{{\mathbb R}}

\newcommand{\e}{{\rm e}}
\newcommand{\E}{{\mathcal E}}
\newcommand{\F}{{\mathcal F}}

\newcommand{\C}{\mathbb{C}}

\def\uno{{\kern+.3em {\rm 1} \kern -.22em {\rm l}}}

\def\norma#1{\left\| #1\right\|}
\def\norm#1{\left\| #1\right\|}

\def\d{{\rm d}}
\def\e{{\rm e}}

\def\Br{B_{\R}}
\def\reg{\sigma}

\numberwithin{equation}{section}

\def\phi{{\varphi}}
\newcommand{\equal}{\buildrel {\rm def} \over {=} }

\begin{document}

\title{Some analytic results on the FPU paradox}
\author{D.~Bambusi, A.~Carati, A.~Maiocchi, A.~Maspero}

\date{\today}
\maketitle

\begin{abstract}
We present some analytic results aiming at explaining the lack of
thermalization observed by Fermi Pasta and Ulam in their celebrated
numerical experiment. In particular we focus on results which persist
as the number $N$ of particles tends to infinity. After recalling the
FPU experiment and some classical heuristic ideas that have been used
for its explanation, we concentrate on more recent rigorous results
which are based on the use of (i) canonical perturbation theory and
KdV equation, (ii) Toda lattice, (iii) a new approach based on the
construction of functions which are adiabatic invariants with large
probability in the Gibbs measure.
\end{abstract}

\section{Introduction}\label{intro}

In their celebrated numerical experiment Fermi Pasta and Ulam
\cite{FPU}, being interested in the problem of foundation of
statistical mechanics, studied the dynamics of a chain of nonlinear
oscillators. In particular they studied the evolution of the energies
of the normal modes and their time averages. FPU considered initial
data with all the energy in the first Fourier mode and observed that
(1) the harmonic energies seem to have a recurrent behaviour (2) the
time averages of the harmonic energies quickly relaxes to a
distribution which is exponentially decreasing with the wave number
(FPU packet of modes). This was quite surprising since, from the
principles of statistical mechanics, the solution was expected to
explore the whole phase space and the energies of the normal modes
were expected to relax to equipartition.

Subsequent numerical and analytic investigations tackled the problem
of understanding such a behaviour and of understanding whether or not
it persists as the number $N$ of the particles tends to infinity. In
particular the interesting regime is that of the thermodynamic limit
in which the specific energy is kept fixed while $N\to\infty$. Indeed,
in order to be relevant for the foundation of statistical mechanics
the FPU paradox (namely the phenomena described above) has to persist
in such a limit.

The aim of the present paper is to present a short review of the
status of the research, focusing only on analytic results and in
particular to a couple of results recently obtained by the authors
\cite{BM14,MBC14}.

The paper is organized as follows: in sect. \ref{num} we recall the
FPU numerical results (we add only one further very old numerical
result showing the existence of a threshold for thermalization). In
sect. \ref{2} we will discuss some theoretical ideas which have been
used in order to try to explain and to understand the FPU paradox. In
particular we will discuss (1) the relation between FPU lattice and
KdV equation, (2) the use of KAM theory and canonical perturbation
theory (and Nekhoroshev's theorem) in the context of
FPU dynamics.  In Sect.~\ref{3} we present some rigorous results that
have been obtained in the last ten years on the problem.  In
Sect.\ref{toda} we present a recent result which exploits the
vicinity of the Toda lattice and the FPU chain in order to
improve known results of the lifetime of the FPU packet. Finally, in
Sect.\ref{thermodynamic} we will present an averaging theorem for the FPU
chain valid in the thermodynamic limit. This last result in particular
deals with a slightly different problem, namely the exchange of
energy among the different degrees of freedom when one starts with an
initial datum belonging to a set of large Gibbs measure. We conclude
the paper with a short discussion Sect. \ref{fine}.

\section{Introduction to FPU paradox}
\label{num}

The Hamiltonian of the FPU--chain can be
written, in suitably rescaled variables, as 
\begin{equation}
\label{H}
H_{FPU}=H_0+H_1+H_2
\end{equation}
where
\begin{eqnarray*}
H_0&\equal &\sum_{j}\left(\frac{p_j^2}2+\frac{\left(q_{j+1}-
  q_j\right)^2}2
\right)\ ,\\
H_1&\equal& \frac 1{3!}\sum_{j}\left(
q_{j+1}- q_j\right)^3\\
H_2&\equal& \frac A{4!}\sum_{j} \left( q_{j+1}- q_j\right)^4\ ,
\end{eqnarray*}
where $(p,q) $ are canonically conjugated variables. We will consider
the case of periodic boundary conditions, i.e.  $q_{-N-1}=q_{N+1}$ and
$p_{-N-1}=p_{N+1}$. 

In order to introduce the Fourier basis consider the vectors
\begin{equation}
\label{2.basis}
{
{\hat e}_{k}(j)=
{\hat e}_{k}(j)=
\begin{cases}
\frac1{\sqrt{N+1}}\sin{\left(\frac{jk\pi}{N+1}\right)},\qquad
k=1,\ldots,N,\cr
\frac1{\sqrt{N+1}}\cos{\left(\frac{jk\pi}{N+1}\right)},\qquad
k=-1,\ldots,-N,\cr
\frac1{\sqrt{2N+2}},\qquad\qquad\qquad k=0,\cr
\frac{(-1)^j}{\sqrt{2N+2}},\qquad\qquad\qquad k=-N-1.\cr
\end{cases}
}
\end{equation}
Unless specifically needed, we will not specify the set where the
indexes $j,$ and $k$ vary.

Introducing the Fourier variables $(\hat p_k,\hat q_k)$ by
\begin{equation}
\label{2.nm}
p_{ j}=\sum_{ k}{\hat p}_{ k}{\hat e}_{ k}({ j})\
  ,\qquad q_{ j}=\sum_{ k}{{\hat q}_{ k}}{\hat
  e}_{ k}({ j})\
\end{equation}
with
\begin{equation}
\label{4}
\omega_k=2\sin\left(\frac{|k|\pi}{2(N+1)}\right)\ .
\end{equation}
the system takes the form
\begin{equation}
\label{2}
H=H_0+H_1+H_2 
\end{equation}
where 
\begin{equation}
\label{3}
H_0(\hat p, \hat q)= \sum_{k} \frac{\hat p_k^2+\omega_k^2\hat
  q_k^2}{2} \ ,\quad H_1=H_1(\hat q) \ ,\quad H_2=H_2(\hat q)\ .
\end{equation}
We also introduce the harmonic energies 
$$ E_k=\frac{\hat p_k^2+\omega_k^2\hat q_k^2}{2} ,
$$
and their time averages
\begin{equation}
\label{5.1}
\langle E_k\rangle (T):=\frac{1}{T}\int_0^TE_k(t)dt\ .
\end{equation}

We will often use also the {\it specific harmonic energies} defined by
\begin{equation}
\label{specific}
\E_k:=\frac{E_k}{N}\ .
\end{equation}

We recall that according to the principles of classical statistical
mechanics, at equilibrium, each of the harmonic oscillators should
have an energy equal to $\beta^{-1}$, $\beta=(k_bT)^{-1}$ being the
standard parameter entering in the Gibbs measure (and $k_b$ being the
Boltzmann constant). Furthermore, if the system has good statistical
properties, the time averages of the different quantities should
quickly relax to their equilibrium value.  

\begin{figure}[ht]
\begin{center}
\includegraphics[width=.7 \textwidth]{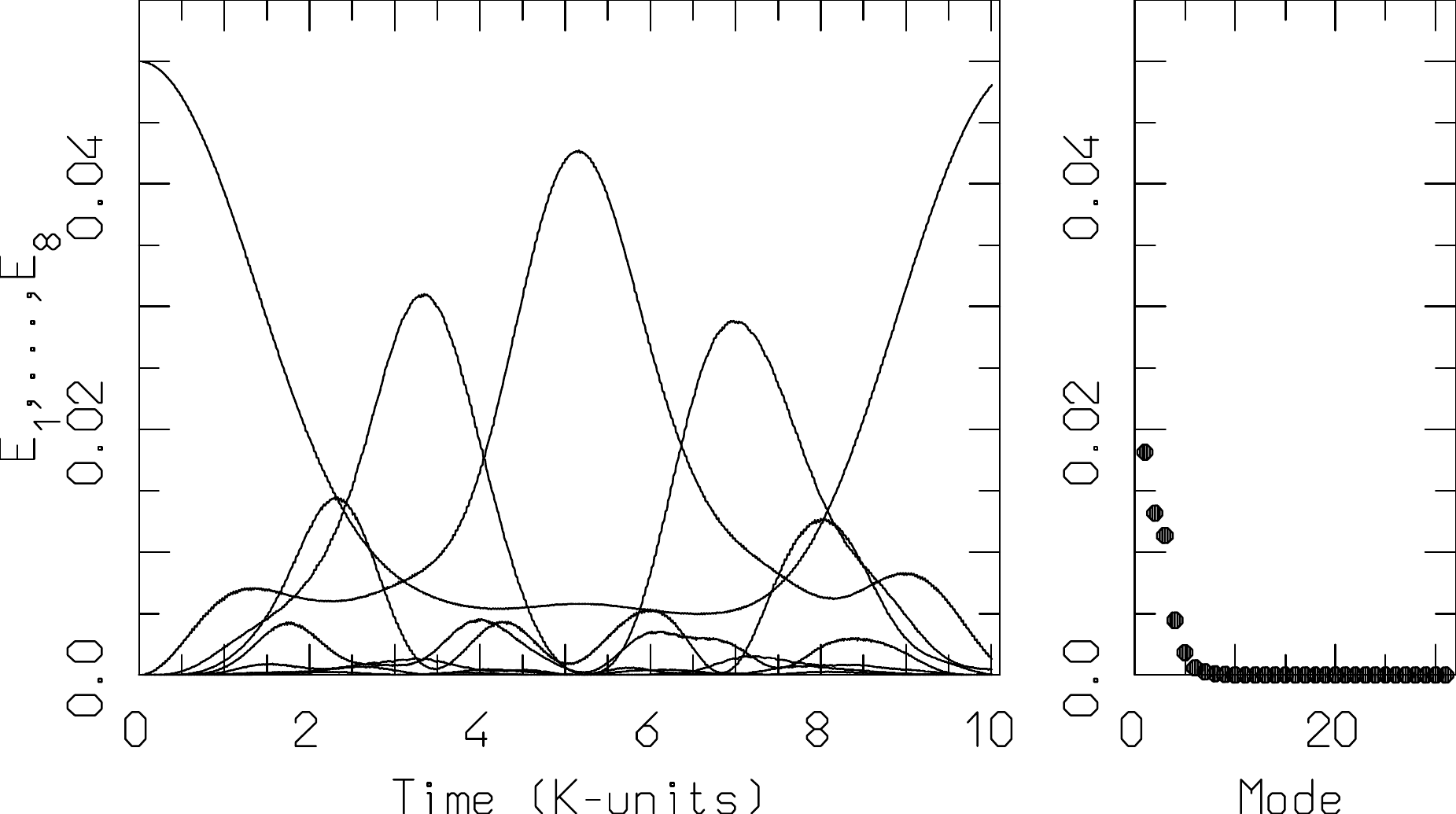}
\caption{Energy per mode and final value of their time averages.}
\label{fig1a}
\end{center}
\end{figure}

Fermi Pasta and Ulam studied the time evolution\footnote{Actually FPU
  studied the case of Dirichlet boundary conditions, but as is well
  known, such a case can be considered as a subcase of that of
  periodic boundary conditions.} of $E_k$ and of $\langle E_k\rangle
$.  Figure \ref{fig1a} shows the results of the numerical computations
by FPU; the initial data are chosen with $E_1(0)\not=0$ and $E_k(0)=0$
for any $|k|>1$.

From Figure \ref{fig1a} one sees that the energy flows quickly
to some modes of low frequency, but after a short period it returns
almost completely to the first mode, in the right part of the figure
the final values of $\langle E_k\rangle (t)$ are plotted in a linear
scale. The final distribution turns out to be exponentially
decreasing with $k$.

\begin{figure}[ht]
\begin{center}
 \makebox[8.5 truecm]{
\includegraphics[width=.6 \textwidth]{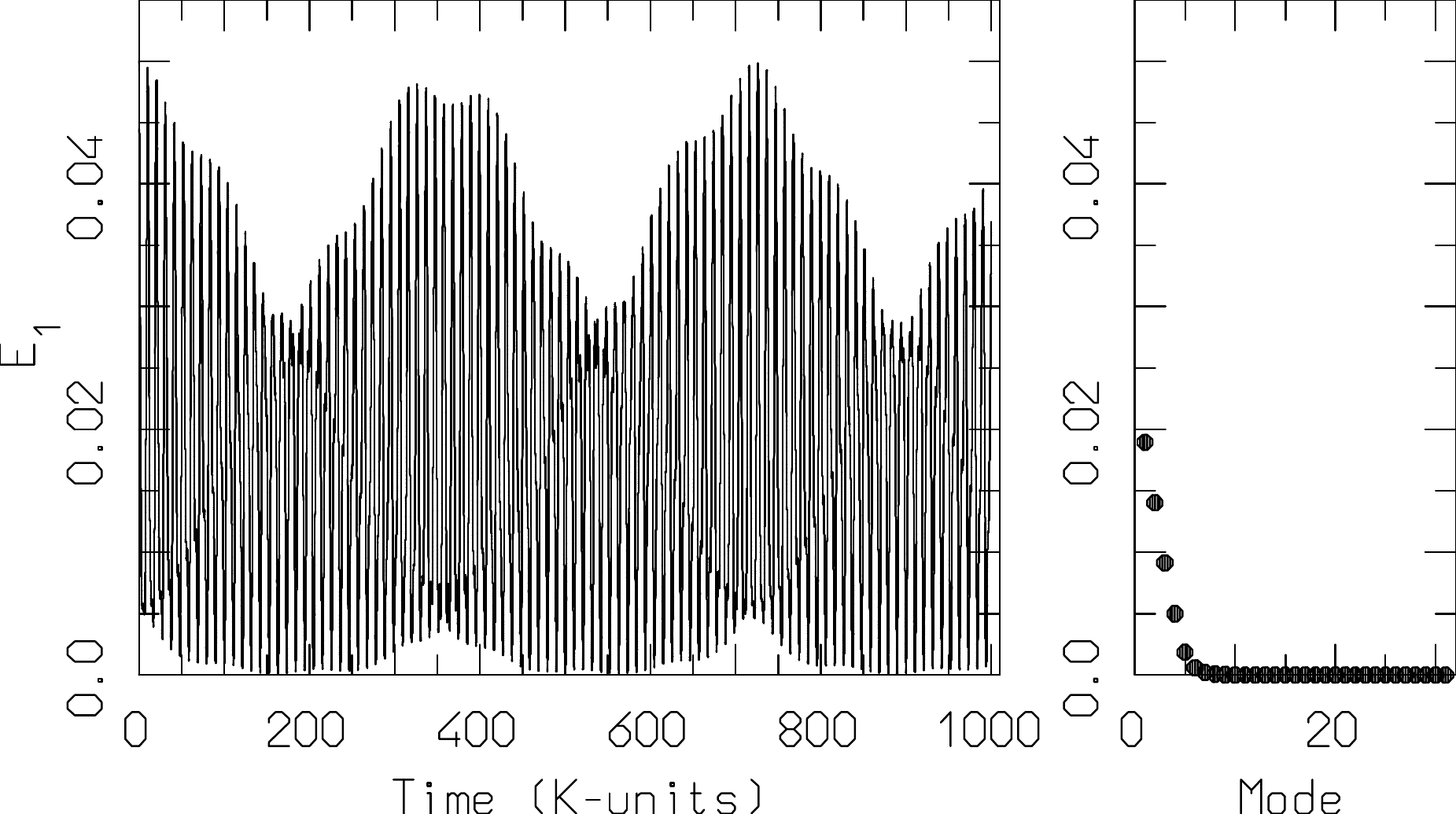}
}
\caption{Energy of the first mode and final value of $\langle
  E_k\rangle(t)$ at longer time scales.}
\label{fig1c}
\end{center}
\end{figure}

If one continues the integration one sees that the phenomenon repeats
almost identically for a very long time (see figure \ref{fig1c}).

\begin{figure}[ht]
\begin{center}
 \makebox[8.5 truecm]{
\includegraphics[width=.6 \textwidth]{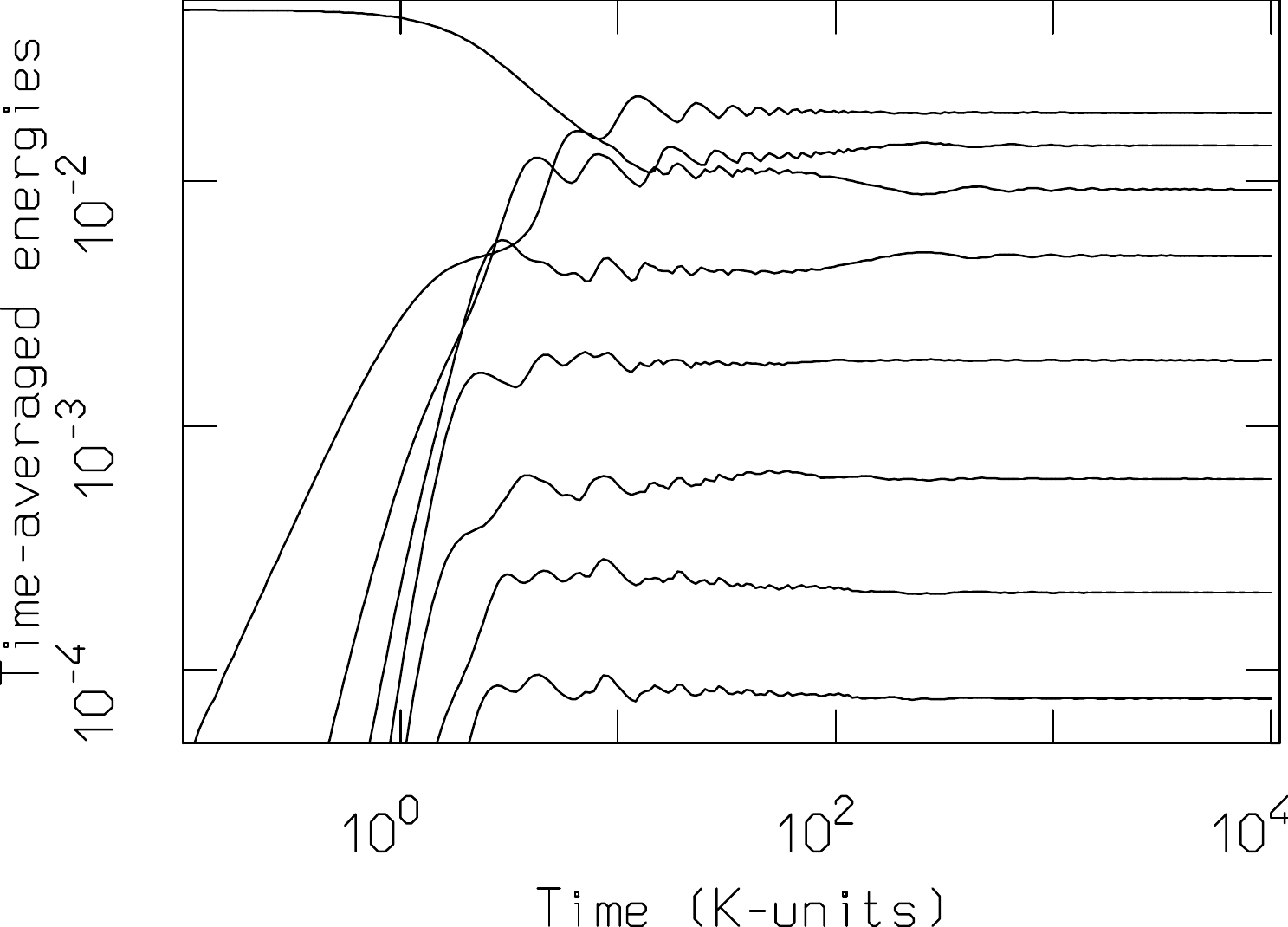}
}
\caption{$\langle E_k\rangle(t)$ versus time.}
\label{fig3}
\end{center}
\end{figure}
	
In figure \ref{fig3} the averages $\langle E_k\rangle
(t)$ are plotted versus time in a semi-log scale. Figure \ref{fig3}
corresponds to initial data with small energy, and one sees that the
quantities 
$\langle E_k\rangle (t)$ quickly relax to well defined values, say
$\bar E_k$. Such values depend on $k$, and, as shown by figure
\ref{fig1c}, decay exponentially.

To describe the situation with the words by Fermi Pasta and Ulam ``The
result shows very little, if any, tendency towards equipartition of
energy among the degrees of freedom.''  This is what is usually known
as the Fermi Pasta Ulam paradox.

\begin{figure}[ht]
\begin{center}
 \makebox[8.5 truecm]{
\includegraphics[width=.6 \textwidth]{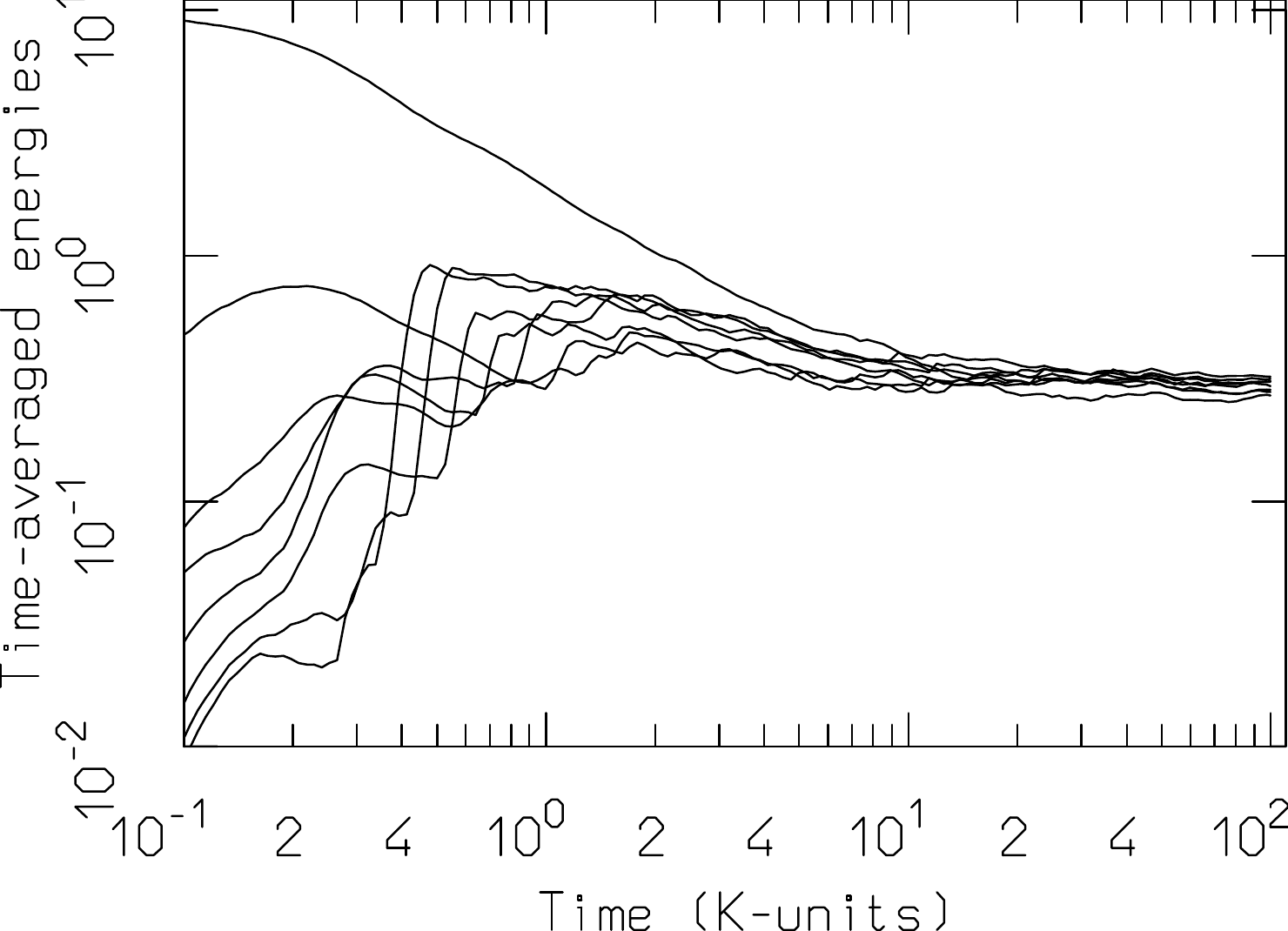}
}
\caption{$\langle E_k\rangle(t)$ versus time at large energy.}
\label{fig4}
\end{center}
\end{figure}

It is interesting to investigate the behaviour of the system when the
energy per particle is increased. This is described in the second of
figures \ref{fig4} from which one sees that the FPU paradox disappears
in this regime: here equipartition is quickly reached.

FPU numerical experiment has originated a huge amount of scientific
research and in particular subsequent numerical computations have
established the shape of the packet of modes to which energy flows (see
e.g. \cite{BGG04}) and have put into evidence that the FPU packet
is only metastable \cite{parisi}, namely that after a quite long time,
whose precise length is not yet precisely established, the system
relaxes to equipartition (see e.g. \cite{BGP04,benettin_ponno}).

\section{Theoretical analysis}\label{2.1} 

We remark that the {\it theoretical} understanding of the FPU
paradox would be absolutely fundamental: indeed it is clear that
the phenomenon has some relevance for the foundation of statistical
mechanics if it persists in the thermodynamic limit, i.e. in the limit
in which the number of particles $N\to\infty$ while the energy per
mode, namely $\sum_{k}E_k/N$ is kept fixed. Of course numerics can
give some indications, but a definitive result can only come from a
theoretical result, which is the only one able to reach the limit
$N=\infty$.

\subsection{KdV}\label{KdV}

One of the first attempts to explain the FPU paradox has been on the
use of the Kortweg de Vries equation (KdV). The point is that on the
one hand KdV is known to approximates the FPU and on the other one KdV
is also known to be integrable, so that it displays (in suitable
variables) a recurrent behaviour.

We now recall briefly the way KdV is introduced as a modulation
equation for the FPU. We also restrict to the subspace
\begin{equation}
\label{media}
\sum_jq_j=0=\sum_{j}p_j
\end{equation}
which is invariant under the dynamics. The idea is to consider
initial data with long wave and small amplitude, namely to interpolate
the difference $q_j-q_{j+1}$ through a smooth small function slowly
changing in space (and time). This is obtained through an Ansatz of
the form
\begin{equation}
\label{interpol}
q_j-q_{j+1}=\epsilon u(\mu j,t), \quad \mu:=\frac{1}{N}\ , \quad
\epsilon\ll 1
\end{equation}
with $u$ periodic of period $2$. It turns out that in
order to fulfill the FPU equations, the function $u $ should have the
form
$$
u(x,t)= f(x-t,\mu^3t )+g(x+t,\mu^3t)\ 
$$
with $f(y,\tau)$ and $g(y,\tau)$ fulfilling the equations 
\begin{equation}
\label{eqkdv}
f_\tau+\frac{\mu^2}{\epsilon}f_{yyy}+ff_y=O(\mu^{2})\ ,\quad
g_\tau-\frac{\mu^2}{\epsilon}g_{yyy}-gg_y=O(\mu^{2}) \ ,
\end{equation}
namely, up to higher order corrections, a couple of KdV equations with
dispersion of order $\mu^2/\epsilon$ describe the system. Now, it is
of the year 1965 the celebrated paper by Zabuski and Kruscal on the
dynamics of the KdV equation which was the starting point of soliton
theory and led in particular to the understanding that KdV is
integrable. Thus, the enthusiasm for the discovery of such a beautiful
and important phenomenology, led the idea that also the FPU paradox could
be explained by the fact that the dynamics of the FPU is described in
some limit by an integrable equation.

In order to transform such a heuristic idea into a theorem one should
fill two gaps, the first one consists in showing that in the KdV
equation a phenomenon of the kind of the formation and persistence of
the packet of modes occurs, and the second one consisting in showing
that the solutions of the KdV equation actually describe well the
dynamics of the FPU, namely that the higher order corrections
neglected in \eqref{eqkdv}, are actually small.

Both problems can be solved in the case $\epsilon=\mu^2$, in which the
KdV equation turns out to be the standard one. Indeed the action angle
coordinates for the KdV equation with periodic boundary conditions
have been constructed and studied in detail \cite{kamkdv} and with
their help one can show that if, in the KdV equation one puts all the
energy in the first Fourier mode, then the energy remains forever
localized in an exponentially localized packet of Fourier modes.

However, if one wants to take the limit $N\to\infty$ while keeping
$\epsilon$ fixed (as needed in order to get a results valid in the
thermodynamic limit), one has to study the dispersionless limit of the
KdV equation and very little is known on the behaviour of action angle
variables in this limit, so that the standard theory becomes
inapplicable. Thus we can say that, {\bf in the KdV equation} the
phenomenon of formation and persistence of the packet is not explained
in the limit corresponding to the thermodynamic limit of the FPU
lattice.

The second problem is also far from trivial, since the perturbation
terms of \eqref{eqkdv} contain higher order derivatives, so we are
dealing with a singular perturbation of KdV and the proof of theorems
connecting the solutions of KdV and the solutions of FPU have only
recently been obtained \cite{SW00,BamPon}. 

\subsection{KAM theory and canonical perturbation theory}\label{KAM} 

Izrailiev and Chirikov \cite{IC} in 1966 suggested to explain the
behaviour observed by FPU through KAM theory. We recall that KAM
theory deals with perturbations of integrable systems and ensures
that, provided the perturbation is small enough, most of the invariant
tori in which the phase space of the unperturbed system is foliated
persist in the complete system. In the case of FPU of course the
integrable system is the linearized chain and the perturbation is
provided by the nonlinearity, so the size of the nonlinearity increases
with the energy of the initial datum and KAM theory should apply for
energy smaller then some $N$-dependent threshold $\epsilon_N$. This
approach has the remarkable feature of potentially explaining the
recurrent behaviour observed by FPU and also the fact that it
disappears for large energy. 

From the argument of Izrailiev and Chirikov (based on Chirikov's
criterion of overlapping of resonances) one can extract also an
explicit estimate of the threshold which should go to zero like
$N^{-4}\equiv \mu^4$. Such an estimate is derived by Izrailiev and
Chirikov by considering initial data on high frequency Fourier mode,
while they do not deduce any explicit estimate for the case of initial
data on low frequency modes. Their argument has been extended to
initial data on low frequency Fourier modes by Shepeliansky
\cite{Shep} leading to the claim that also corresponding to such kind
of initial data FPU phenomenon should disappear as
$N\to\infty$, however a subsequent reanalysis of the problem has led
to different conclusions \cite{ponno_cargese}, so, at least, we can
say that the situation is not yet clear.

We emphasize that the actual application of KAM theory to the FPU
lattice is quite delicate since the hypotheses of KAM theory involve a
Diophantine type nonresonance condition and also a nondegeneracy
condition. The two conditions have been verified only much later by
Rink \cite{Rin01} (see also \cite{nishida,HK1}). Then one has to
estimate the dependence of the threshold $\epsilon_N$ on $N$ and it
turns out that a rough estimate gives that $\epsilon_N$ goes to zero
exponentially with $N$ (essentially due to the Diophantine type
nonresonance condition).

In order to weaken this condition on $\epsilon_N$, Benettin, Galgani,
Giorgilli and collaborators
\cite{rotatori1,rotatori2,vincoli1,vincoli2,martinoli,catene} started
to investigate the possibility of using averaging theory and
Nekhoroshev's theorem to explain the FPU paradox. This a quite remarkable
change of point of view, since averaging theory and Nekhoroshev's
theorem give results controlling the dynamics over long, but finite
times, so such a point of view leaves open the possibility that the
FPU paradox disappears after a long but finite time, which is what is
actually seen in numerical investigations (see also the remarkable
theoretical paper \cite{parisi}). Results along this line have been
obtained for chains of rotators (\cite{rotatori1,catene}) and FPU
chains with alternate masses \cite{martinoli,catene}. An application
to the true FPU model is given in the next section.

\section{Some rigorous results}\label{3.1}

\subsection{KdV and FPU}\label{metafpu}

The unification of the two points of view above has been obtained in
the paper \cite{BamPon}, in which canonical perturbation theory has
been used in order to deduce a couple of KdV equation playing the role
of resonant normal form for the FPU lattice and this has been used in
order to describe the phenomenon of formation and metastability of the
FPU packet.  We briefly recall the result of \cite{BamPon}. 

We consider here the case of periodic boundary conditions. Consider a
state of the form \eqref{interpol} and write the equation for
the evolution of the function $u$, then it turns out that such an
equation is a Hamiltonian perturbation of the wave equation, so one
can use canonical perturbation theory for PDEs in order to simplify
the equation. Passing to the variables $f,g$ the normal form turns out
to be the Hamiltonian of a couple of non interacting KdV equations. In
\cite{BamPon} a rigorous theory estimating the error was developed,
and the main results of that paper are contained in Theorem
\ref{s.200} and Corollary \ref{metaf} below.

Consider the KdV equation 
$$
f_\tau+f_{yyy}+ff_y=0\ ,
$$ it is well known \cite{kamkdv} that if the initial datum extends
to a function analytic in a complex strip of width $\sigma$, then the
solution (as a function of the space variable $y$) is also analytic
(in general in a smaller complex strip). 

Consider now a couple of solutions $f,g$ of KdV with analytic initial
data and 
let $q^{KdV}_j(t)$ be the unique sequence such that
\begin{align}
\label{solKdV}
&q_j^{KdV}(t)-q_{j+1}^{KdV}(t)=\mu^2\left[f\left(\mu(j-t),\mu^3t\right)
  +g\left(-\mu(j+t),\mu^3t\right) 
  \right]\ ,
\\
\nonumber
 &\sum_{j}q_j^{KdV}(t)\equiv 0\ ,
\end{align}
where, as above, $\mu:=N^{-1}$. Then the result is that $q_j^{KdV}$
approximates well the true solution of the FPU lattice.

Let $q_j(t)$ be the solution of the FPU equations with the
initial data $q_j(0)=q_j^{KdV}(0)$, $\dot q_j(0)=\dot q_j^{KdV}(0)$;
denote by $E_k(t)$ the energy in the $k^{th}$ Fourier mode of the
solution of the FPU with such initial datum and $\E_k:=E_k/N$.

The following theorem holds
\begin{theorem}
\label{s.200}\cite{BamPon}
Fix an arbitrary $T_f>0$. Then there exists $\mu_*$ such that, if
$\mu<\mu_*$ then for all times $t$ fulfilling
\begin{equation}
\label{s.204}
\left|t\right|\leq\frac{T_f}{\mu^3}
\end{equation}
one has 
\begin{equation}
\label{s.203}
\sup_{j}\left|r_j(t)-r^{KdV}_j(t)\right|\leq C\mu^3\ \ , 
\end{equation}
where $r_j:=q_{j}-q_{j+1}$ and similarly for $r_j^{KdV}$.
Furthermore, there exists $\sigma>0$ s.t., for the same times, one has
\begin{equation}
\label{sti.fourKdV}
{\E_k(t)}\leq C\mu^4e^{-\sigma |k|}+C\mu^5\ .
\end{equation}
\end{theorem}
Exploiting known results on the dynamics of KdV (and Hill's
operators \cite{poschel}) one gets the following corollary which is directly
relevant to the FPU paradox.
\begin{corollary}
\label{metaf}
Fix a positive $R$ and a positive $T_f$, then there exists a positive
constant $\mu_*$, with the following property: assume $\mu<\mu_*$ and
consider the FPU system with an initial datum fulfilling
\begin{equation}
\label{M.1.11}
\E_{1}(0)=\E_{-1}(0)=R^2 \mu^4\ \ ,\quad
\E_k(0)\equiv \E_k(t)\big\vert_{t=0}=0\ ,\quad \forall
|k|\not=1,\ .
\end{equation}
  Then, along the corresponding solution, equation \eqref{sti.fourKdV}
  holds for the times \eqref{s.204}.

Furthermore there exists a sequence of almost periodic functions
  $\{F_k\}$ such that, defining the specific energy
  distribution
\begin{equation}
\label{limd}
 \F_{k}= \mu^4 F_k\ ,
\end{equation}
one has
\begin{equation}
\label{A.1.3}
\left|\E_{k}(t)- \F_k(t) \right|
\leq C_2\mu^5 \ ,\quad 
\left|t\right|\leq \frac{T_f}{\mu^3}\ .
\end{equation}
\end{corollary}

\begin{remark}
\label{A12}
One can show that the following limit exists
\begin{equation}
\label{mediak}
\bar F_k :=\lim_{T\to\infty}\frac{1}{T}\int_0^TF_k(t)dt\ .
\end{equation}
It follows that up to a small error the time average of $\E_{k}(t)$
relaxes to the limit distribution obtained by rescaling $\bar F_k$. Of
course $\bar F_k$ is exponentially decreasing with $k$, but one can
also show that actually one has $\bar F_k\not=0$ $\forall k\not=0$
\end{remark}

The strong limitation of the above results rests in the fact that they
only apply to initial data with specific energy of order $\mu^4$, thus
they do not apply to the thermodynamic limit.

\subsection{Longer time scales with less energy}\label{tedeschi}

We present here a result by Hairer and Lubich \cite{HL12} which is
valid in a regime of specific energy smaller then that considered
above, but controls the dynamics for longer time scales. The proof of
the result is based on the technique of modulated Fourier expansion
developed by the authors and collaborators. In some sense such a
technique can be considered as a variant of classical perturbation
theory. The key tool that they use for the proof is an accurate
analysis of the small denominators entering in the perturbative
construction.

To be precise \cite{HL12} deals with the case of periodic boundary
conditions. 

\begin{theorem}
\label{HLt}
There exist positive constants $R_*$, $N_*$, $T$, with the following
property: consider the FPU system with an initial datum fulfilling
\eqref{M.1.11} {\bf with $R<R_* $}.  Then, along the corresponding
solution, one has
\begin{equation}
\label{M.1.21}
\E_k(t)\leq R^2 \mu^4 R^{2(|k|-1)}\ ,\quad \forall \, 1\leq | k| \leq  N  \ ,\quad
\forall |t|\leq\frac{T}{\mu^2 R^5} \ .
\end{equation}
\end{theorem}

It is interesting to compare the time scale covered by this theorem
with the time scale of Corollary \ref{metaf}. It is clear that the
time scale \eqref{M.1.21} is longer than \eqref{s.204} as far as
\begin{equation}
\label{hl+meta}
R<N^{-1/5}\ 
\end{equation}
(where we made the choice $T_f:=T$), namely in a regime where the
specific energy goes to zero faster then in the Theorem \ref{s.200}. 

One has also to remark that in Theorem \ref{HLt} one gets an
exponential decay of the Fourier modes valid for all $k$'s (the term
of order $\mu^5$ present in \eqref{sti.fourKdV} is here absent).

\section{Toda lattice}\label{s.toda}

It is well known that close to the FPU lattice there exists a remarkable
integrable system, namely the Toda lattice \cite{toda,henon} whose
Hamiltonian is given by
 \begin{equation}
H_{Toda}(p,q)=\frac{1}{2} \sum_j{p_j^2} + \sum_j
e^{q_j-q_{j+1}} \ , 
\label{toda}
\end{equation}
(we consider the case of periodic boundary conditions), so that
 one has
$$H_{FPU}(p,q) = H_{Toda}(p,q) + (A-1)H_2(q) + H^{(3)}(q), $$ 
where
\begin{align*}
H_l( q) &:= \sum_{j} \frac{(q_j - q_{j+1})^{l+2}}{(l+2)!}\ ,\quad \forall
l\geq2\ ,
\\
H^{(3)}&:= -\sum_{l \geq 3} H_l\ ,
\end{align*}
which shows the vicinity of $H_{FPU}$ and $H_{Toda}$.

The idea of exploiting the Toda lattice in order to deduce
information on the dynamics of the FPU chain is an old one; however
in order to make it effective, one has first to deduce information on
the dynamics of the Toda lattice itself, and this is far from trivial. The
most obvious way to proceed consists in constructing action angle
coordinates for the Toda lattice and using them to study the
dynamics. An important result in this program was obtained by Henrici
and Kappeler \cite{HK0,HK1} who constructed action angle coordinates
and Birkhoff coordinates (a kind of cartesian action angle
coordinates) showing that, for any $N$, such coordinates are globally
analytic (see Theorem \ref{HK} below for a precise statement). However
the construction by Henrici and Kappeler is not uniform in the number
of particles $N$, thus it is not possible to exploit it directly in
order to get results for the FPU paradox in the limit $N\to\infty$.

Results on the behaviour of the integrable structure of Toda for large
$N$ have been recently obtained in a series of papers
\cite{thierrygolse,BKP0,BKP1,BKP2,BM14}. In particular in
\cite{BKP0,BKP1,BKP2}, exploiting ideas from \cite{thierrygolse}, it has been
shown that as $N\to\infty$ the actions and the frequencies of the Toda
lattice are well described by the actions and the frequencies of a
couple of KdV equations, at least in a regime equal to that of Theorem
\ref{s.200}, namely of specific energy of order $\mu^4$.

Further results (exploiting some ideas from \cite{BKP1,BKP2,BKP0}) directly
applicable to the FPU metastability problem have been obtained in
\cite{BM14} and now we are going to present them. In \cite{BM14} the
regularity properties of the Birkhoff map, namely the map introducing
Birkhoff coordinates for the FPU lattice, have been studied and lower
and upper bounds to the radius of the ball over which such a map is
analytic have been given.

\vskip20pt

To come to a precise statement we start by recalling the result by
Henrici and Kappeler. 

Consider the Toda lattice in the submanifold \eqref{media} and
introduce the linear Birkhoff variables
\begin{equation}
\label{lin.bir.real}
X_k=\frac{\hat p_{k}
}{\sqrt{\omega_k} }\ ,\quad  Y_k={\sqrt{\omega_k} }{\hat q_k 
}\ ,\quad |k|=1,...,N
\end{equation}
using such coordinates, $H_0$ takes the form
\begin{equation}
\label{quad.part}
H_0
= \sum_{|k|=1}^{N}\omega_k \frac{X_k^2+Y_k^2}{2}\ .
\end{equation}
With an abuse of notations, we re-denote by $H_{Toda}$ the Hamiltonian
\eqref{toda} written in the coordinates $(X,Y)$. 

\begin{theorem}[{{\cite{kapphen2}}}]
\label{HK}For any integer $N \geq 2$ there exists a global real
analytic canonical diffeomorphism 
$\Phi_N:\R^{2N}\times\R^{2N} \to \R^{2N}\times\R^{2N}$, $(X,
Y)=\Phi_N(x,y)$ with the following properties:
\begin{itemize}
\item[(i)] The Hamiltonian $H_{Toda}\circ \Phi_N$ is a function of the
  actions $I_k:=\frac{x^2_k+y_k^2}{2}$ only, i.e. $(x_k,y_k)$ are 
  Birkhoff variables for the Toda Lattice.
\item[(ii)] The differential at the origin is the identity:
  $d\Phi_N(0,0)=\uno$. 
\end{itemize}
\end{theorem}

In order to state the analyticity properties fulfilled by the map
$\Phi_N$ as $N\to\infty$ we need to introduce suitable norms: for any
$\sigma\geq0$ define
\begin{equation}
\norma{(X,Y)}^2_{{\reg}}:=\frac{1}{N}\sum_{k
  }\, e^{2\sigma |k|}\,\omega_{k}\,
\frac{\left|X_k\right|^2 + \left|Y_k\right|^2}{2}
\label{nor.bir}
\end{equation}
We denote 
by $B^{{\reg}}(R)$  the ball in $\C^{2N}\times \C^{2N}$ of
radius $R$ and center $0$ in the topology defined by the norm
$\norm{.}_{{\reg}}$. We will also denote by
$\Br^{{\reg}}:=B^{{\reg}}(R)\cap(\R^{2N}\times\R^{2N})$ the {\it real}
ball of radius $R$.
\begin{remark}
\label{sullenorme}
We are particularly interested in the case
$\sigma>0$ since, in such a case, states with finite norm
are exponentially decreasing in Fourier space. 
\end{remark}

The main result of \cite{BM14} is the following Theorem.

\begin{theorem}
\label{main}\cite{BM14}
Fix $\sigma\geq 0$ then there exist $R,R'>0$ s.t. $\Phi_N$ is analytic
on $B^\sigma\left(\frac{R}{N^\alpha}\right)$ and fulfills
\begin{equation}
\label{analytic}
\Phi_N\left( B^\sigma\left(\frac{R}{N^\alpha}\right) \right) \subset
B^\sigma\left(\frac{R'}{N^\alpha}\right)\ ,\quad \forall N\geq 2
\end{equation}
if and only if $\alpha\geq 2$. The same is true for the
inverse map $\Phi_{N}^{-1}$.
\end{theorem}

\begin{remark}
\label{size}
A state $(X,Y)$ is in the ball $B^{{\reg}}(R/N^2)$ if
and only if there exist interpolating periodic functions
$(\beta,\alpha)$, namely functions s.t.
\begin{equation}
\label{interpol1}
 p_j=\beta\left(\frac{j}{N}\right)\ ,\quad
q_j-q_{j+1}=\alpha\left(\frac{j}{N}\right)\ ,
\end{equation}
which are analytic in a strip of width $\sigma$ and have an analytic
norm of size $R/N^2$. Thus we are in the same regime to which Theorem
\ref{s.200} apply.
\end{remark}

 Theorem \ref{main} shows that the Birkhoff coordinates are analytic
 only in a ball of radius of order $N^{-2}$, which corresponds to
 initial data with specific energy of order $N^{-4}$.

We think this is a strong indication of the fact that standard
integrable techniques cannot be used beyond such regime.

As a corollary of Theorem \ref{main}, one immediately gets that in the
Toda Lattice the FPU metastable packet of modes is actually stable,
namely it persists for infinite times. Precisely one has the following
result.  

\begin{corollary}
\label{M.1}
Consider the Toda lattice \eqref{toda}. Fix $\sigma>0$,
then there exist constants $R_0,$ $C_1,$ such that the following
holds true. Consider an initial datum fulfilling \eqref{M.1.11}
{\bf with $R<R_0 $}.  Then, along the corresponding solution, one has
\begin{equation}
\label{M.1.2}
\E_k(t)\leq R^2(1+ C_1R) \mu^4 e^{-2\sigma |k|}\ ,\quad \forall \,
1\leq |k| \leq N \ ,\quad \forall t\in\R \ .
\end{equation}
\end{corollary}

We recall that this was observed numerically by Benettin and Ponno
\cite{benettin_ponno,benettin_ponno2}. One has to remark that according to the
numerical computations of \cite{benettin_ponno}, the packet exists and
is stable over infinite times also in a regime of finite specific
energy (which would correspond to the case $\alpha=0$ in Theorem
\ref{main}). The understanding of this behaviour in such a regime is
still a completely open problem. 

Concerning the FPU chain, Theorem \ref{main} yields the following
result.
\begin{theorem}
\label{N.1}
Consider the FPU system. Fix $\sigma\geq0$; then there exist constants
$R'_0,$ $C_2,$ $T,$ such that the following holds true. Consider a
real initial datum fulfilling \eqref{M.1.11} {\bf with $R<R'_0$},
then, along the corresponding solution, one has
\begin{equation}
\label{N.1.2}
\E_k(t)\leq {16R^2 \mu^4 e^{-2\sigma |k|}}
\ ,\quad \forall \, 1\leq |k| \leq  N  \ ,\quad
|t| \leq \frac{T}{R^2\mu^4}\cdot\frac{1}{|A-1| + C_2 R\mu^2} \ .
\end{equation}
Furthermore, for $1 \leq |k| \leq N$, consider the action
$I_k:=\frac{x_k^2+y_k^2}{2}$ of the Toda lattice and let $I_k(t)$ be
its evolution according to the FPU flow. Then one has
\begin{equation}
\label{N.1.3}
\frac{1}{N}\sum_{|k| =1}^{N}e^{2\sigma  |k|}\omega_{k}
|{I_k(t) - I_k(0)}|
  \leq C_3 R^2\mu^5 \qquad \mbox{ for } t \mbox{ fulfilling } \eqref{N.1.2}
\end{equation}
\end{theorem}

So this theorem gives a result which covers times one order of
magnitude longer then those covered by Theorem \ref{s.200}.
Furthermore the small parameter controlling the time scale is the
distance between the FPU and the Toda

This is particularly relevant in view of the fact that, according to
theorem \ref{s.200} the time scale of formation of the packet is
$\mu^{-3}$, thus the present theorem shows that the packet persists at
least over a time scale one order of magnitude longer then the time
needed for its formation.

\section{An averaging theorem in the thermodynamic
  limit}\label{thermodynamic} 

In this section we discuss a different approach to the study of the
dynamics of the FPU dynamics, which allows to give some results valid
in the thermodynamic limit. Such a method is a development of the one
introduced in \cite{Car07} in order to deal with a chain of rotators
(see also \cite{RH13}), and developed in \cite{carati_maiocchi} in order to
study a Klein Gordon chain.

We consider here the case of Dirichlet boundary conditions and endow
the phase space by the Gibbs measure at inverse temperature $\beta$,
namely
\begin{equation}
\label{gibbs}
\d \mu(p,q)\equal \frac{\e^{-\beta H_{FPU}(p,q)}}{Z(\beta)} \d p\d q\ ;
\end{equation}
where as usual 
$$Z(\beta):=\int \e^{-\beta H_{FPU}(p,q)}\d p\d q
$$ is the partition function (the integral is over the whole phase
space).  Given a function $F$ on the phase space, we define
\begin{align}
\label{ave}
&\langle F\rangle\equal \int F\d \mu\ ,
\\
\label{norma}
& \norma{F}^2\equal \int  |F|^2\d \mu\ ,
\\
\label{norma1}
& \sigma_F^2\equal \norma{F-\langle F\rangle }^2\ ,
\end{align}
which are called respectively the average, the $L^2$ norm and the
variance of $F$. The correlation of two dynamical variables $F,G$ is
defined by
$$
C_{F,G}:={\langle FG\rangle -\langle F\rangle\langle
  G\rangle} 
$$
and the time autocorrelation of a dynamical variable by 
\begin{equation}
\label{norma2}
C_F(t):=C_{F,F(t)}\ ,
\end{equation}
where $F(t):=F\circ g^t$ and $g^t$ is the flow of the FPU system. 

Remark that the Gibbs measure is asymptotically concentrated on the
energy surface of energy $N/\beta$, thus studying the system in such a
phase space one typically considers data with specific energy equal to
$\beta^{-1}$. 

Let $g\in \mathcal C^2([0,1],\R^+)$ be a twice differentiable
function; we are interested in the time
evolution of quantities of the form
 $$\Phi_g\equal
\sum_{k=1}^N g\left(\frac{k}{N+1}\right) E_k\ .
$$ 

The following theorem was proved in \cite{MBC14}
\begin{theorem}\label{teor:principale}
Let $g\in\mathcal C^2([0,1];\R^+)$ be a function fulfilling
$g'(0)=0$. There exist constants $\beta^*>0$, $ N^*>0$ and $C>0$ s.t.,
for any $\beta>\beta^*$ and for any $N> N^*$, any
$\delta_1,\delta_2>0$ one has
\begin{equation}
\label{prob.1}
P\left(\left|\Phi_g(t)-\Phi_g(0)\right|\geq
\delta_1\sigma_{\Phi_g}\right) \leq \delta_2\ ,\quad |t|\leq \frac{\delta_1\sqrt{\delta_2}}{C}\beta
\end{equation}
where, as above, $\Phi_g(t)=\Phi_g\circ g^t$.
\end{theorem}
This theorem shows that, with large probability, the energy of the
packet of modes with profile defined by the function $g$ remains
constant over a time scale of order $\beta^{-1}$. We also emphasize
that the change in the quantity $\Phi_g$ is small compared to its
variance, which establishes the order of magnitude of the difference
between the biggest and the smallest value of $\Phi_g$ on the energy
surface. 

Theorem \ref{teor:principale} is actually a corollary of a result
controlling the evolution of the time autocorrelation function of
$\Phi_g$. We point out that, in some sense the time autocorrelation
function is a more important object, at least if one is interested in
the problem of dynamical foundation of thermodynamics, indeed, by
Kubo linear response theory the quantity which enters in the
measurements of the specific heat of the chain is exactly the time
autocorrelation function.

\begin{remark}
\label{tanti}
Of course one can repeat the argument for different choices of the
function $g$. For example one can partitions the interval $[0,1]$ of
the variable $k/(N+1)$ in $K$ sub-intervals and define $K$ different
functions $g^{(1)}, g^{(2)},..., g^{(K)} $, with disjoint support,
each one fulfilling the assumptions of Theorem \ref{teor:principale},
so that one gets that the quantities $\Phi_{g^{(l)}}\equal \sum_k
g_k^{(l)}E_k $ are adiabatic invariants, i.e. the energy essentially
does not move from one packet to another one.
\end{remark}

\vskip10pt
The scheme of the proof of Theorem \ref{teor:principale} is as
follows: first, following ideas coming from celestial mechanics, one
performs a formal construction of an integral of motion as a power
series in the phase space variables. As usual, already at the first
step one has to solve the so called homological equation in order to
find the third order correction of the quadratic integral of
motion. The solution of such an equation involves some small
denominators which are usually the source of one of the problems
arising when one wants to control the behaviour of the system in the
thermodynamical limit. Here we show that, if one takes as the
quadratic part of the integral the quantity $\Phi_g$, then every small
denominator appears with a numerator which is also small, so that the
ratio is bounded. The subsequent step consists in adding rigorous
estimates on the variance of the time derivative of the so constructed
approximate integral of motion. This allows to conclude the proof. 

We emphasize that this procedure completely avoids to impose the time
invariance of the domain in which the theory is developed, which is
the requirement that usually prevents the applicability of canonical
perturbation theory to systems in the thermodynamic limit.  Indeed in
the probabilistic framework the relevant estimates are global in the
phase space.

\section{Conclusions}\label{fine}

Summarizing the above results, we can say that all the analytic
results available nowadays can be split into two groups: the first
group consisting of those which describe the formation of the packet
observed by FPU and give some estimates on its time of persistence.
Such results do not survive in the thermodynamic limit; instead they
are all confined to the regime in which the specific energy is order
$N^{-4}$. We find particularly surprising the fact that very different
methods lead to the same regime and of course this raises the suspect
that there is some reality in this limitation. However one has to say
that numerics do not provide any evidence of changes in the dynamics
when energy is increased beyond this limit.

A few more comments on this point are the following ones: the
limitation appearing in constructing the Birkhoff variables in Toda
lattice (which are the source of the limitations in the applicability
of Theorem \ref{N.1}) are related to the fact that one is implicitly
looking for an integral behaviour of the system, namely a behaviour in
which the system is essentially decoupled into non interacting
oscillators. On the contrary the construction leading to Theorem
\ref{s.200} is based on a resonant perturbative construction in which
the small denominators are not present. The main
limitation for the applicability of Theorem \ref{s.200} comes from the
need of considering the zero dispersion limit of the KdV equation.
So, it is surprising that the regime at which the two results apply is
equal.

So the question on whether the phenomenon of formation of a metastable
packet persists in the thermodynamic limit or not is still completely
open. An even more open question is that of the length of the time
interval over which it persists. Up to now the best result we know is
that of Theorem \ref{N.1}, but, from the numerical experiments one
would expect longer time scales (furthermore in the thermodynamic
limit). How to reach them is by now not known.

\vskip 10pt

At present the only known result valid in the thermodynamic limit is
that of Theorem \ref{teor:principale}. However we think that this
should be considered only as a preliminary one. Indeed it leaves open
many important questions. The first one is the optimality of the time
scale of validity: the technique used for its proof does not extended
to higher order construction. This is due to the fact that at order
four new kind of small denominators appear and up to now we have not
been able to control them. Furthermore there is no numerical evidence
of the optimality of the time scale controlled by such theorem.

An even more important question is that of the relevance of the result
for the foundations of statistical mechanics. Indeed, one expects that
the existence of many integrals of motion independent of the energy
should have some influence on the measurement of thermodynamic
quantities, for example the specific heat. In particular, since the
time needed to exchange energy among different packets of modes
increases as the temperature decreases one would expect that some new
behaviour appears as one lowers the temperature towards the absolute
zero. However up to now we have not been able to put into evidence
some clear effect of the mathematical phenomenon described by Theorem
\ref{teor:principale}. This is one the main goal of our group for the
next future.

\newcommand{\etalchar}[1]{$^{#1}$}
\def\cprime{$'$}
\providecommand{\bysame}{\leavevmode\hbox to3em{\hrulefill}\thinspace}
\providecommand{\MR}{\relax\ifhmode\unskip\space\fi MR }
\providecommand{\MRhref}[2]{%
  \href{http://www.ams.org/mathscinet-getitem?mr=#1}{#2}
}
\providecommand{\href}[2]{#2}

\end{document}